\documentclass[useAMS]{mn2e}
\usepackage{amsmath}
\usepackage{times}
\usepackage{graphicx}

\newcommand{\Teff}{\ensuremath{T_{\rm eff}}}                
\newcommand{\logg}{\ensuremath{\log g}}                     
\newcommand{\Msun}{\ensuremath{\,{\rm M}_\odot}}            
\newcommand{\Rsun}{\ensuremath{\,{\rm R}_\odot}}            
\newcommand{\Lsun}{\ensuremath{\,{\rm L}_\odot}}            
\newcommand{\ion}[2]{{#1}\,{\sc {\small{#2}}}}              
\newcommand{\kms}{\,km\,s$^{-1}$}                           
\newcommand{\cmss}{\,cm\,s$^{-2}$}                          
\newcommand{\micro}{\ensuremath{v_{\rm turb}}}              
\newcommand{\Veq}{\ensuremath{V_{\rm eq}}}                  
\newcommand{\Vsini}{\ensuremath{v \sin i}}                  
\newcommand{\Vsync}{\ensuremath{V_{\rm synch}}}             
\newcommand{\spd}{{\sc spd}}
\newcommand{\mc}[1]{\multicolumn{2}{c}{#1}}

\title[Tracing CNO exposed layers in the binary system u Her]
{Tracing CNO exposed layers in the Algol-type binary system u Her}

\author[V.\ Kolbas et al.]
       {V.~Kolbas$^1$, A.~Dervi\c{s}o\v{g}lu$^2$,  K.~Pavlovski$^1$, and  J.~Southworth$^3$ \\
       $^1$\,Department of Physics, University of Zagreb, Bijeni\v{c}ka cesta 32, 10000 Zagreb, Croatia \\
       $^2$\,Department of Astronomy \& Space Sciences, Erciyes University, Kayseri, Turkey \\
       $^3$\,Astrophysics Group, Keele University, Staffordshire, CV4 7AL, UK \\
}

\date{}

\pagerange{\pageref{firstpage}--\pageref{lastpage}} \pubyear{2014}

\begin{document} \maketitle \label{firstpage}

\begin{abstract}
The chemical composition of stellar photospheres in mass-transferring binary systems is
a precious diagnostic of the nucleosynthesis processes that occur deep within stars,
and preserves information on the components' history.
The binary system u\,Her belongs to a group of hot Algols with both components being B-stars.
 We have isolated the individual spectra of the two components by the technique of spectral
 disentangling of a new series of 43 high-resolution \'echelle spectra. Augmenting these
with an analysis of the {\it Hipparcos} photometry of the system yields revised stellar
quantities for the components of u\,Her. For the primary component (the mass-gaining star)
 we find $M_{\rm A} = 7.88\pm0.26$\Msun, $R_{\rm A} = 4.93\pm 0.15$\Rsun\ and $T_{\rm eff, A}
 = 21\,600\pm220$\,K. For the secondary (the mass-losing star) we find $M_{\rm B} =
2.79\pm0.12$\Msun, $R_{\rm B} = 4.26\pm 0.06$\Rsun\ and $T_{\rm eff, B} = 12\,600\pm550$ K.
A non-LTE analysis of the primary star's atmosphere reveals deviations in the abundances
of nitrogen and carbon from the standard cosmic abundance pattern in accord with
theoretical expectations for CNO nucleosynthesis processing. From a grid of calculated
 evolutionary models the best match to the observed properties of the stars in u\,Her
enabled tracing the initial properties and history of this binary system. We confirm that
it has evolved according to case A mass transfer. A detailed abundance analysis of the
primary  star gives C/N $=$ 0.9, which supports the evolutionary calculations
and indicates strong mixing in the early evolution of the secondary component, which was
originally the more massive of the two.
The composition of the secondary component would be a further
important constraint on the initial properties of u\,Her system, but requires spectra
of a higher signal to noise ratio.
\end{abstract}

\begin{keywords}
stars: fundamental parameters --- stars: binaries: eclipsing --- stars: binaries:
spectroscopic --- stars: individual: u Her
\end{keywords}

\section{Introduction}                                           \label{sec:intro}

The evolution of a star in a binary system is affected by the presence of its companion.
Only a limited space is allowed for evolution due to the mutual gravitational pool of the
components, and the star which was initially more massive will be the first to reach this
limiting radius (i.e.\ the Roche lobe). At this point a rapid phase of mass transfer happens.
 Most of the more massive component is accreted by its companion, and an Algol-type binary
system is formed. The previously more massive star is now a low-mass subgiant filling its
Roche lobe, and its companion is now the hotter and more massive component with the
characteristics of a main
sequence star. The mass-transfer scenario, first hypothesized
by Crawford (1955), is a well-established solution to the Algol paradox (c.f.\ Hilditch 2001).

This evolutionary process causes many observable effects (changes in orbital period,
erratic light variability, distorted radial velocity curves, etc.), but one is particularly
important. Up to 80\% of the mass of the initially more massive star can be lost, exposing
layers which were originally deep within the star and have been altered by thermonuclear
fusion during the star's main sequence evolution. Some of the material transferred to
the companion is similarly altered. The surface chemical compositions of both stars are
therefore a precious diagnostic of the nucleosynthesis processes that occur deep within
stars. The abundance pattern in Algol-type binaries could reveal their past, and would
be strong evidence for postulated mass transfer between the components (c.f.\ Sarna \&
De Greve 1996).

In pioneering studies a general trend has been revealed with an underabundance of carbon
and an overabundance of nitrogen relative to solar values (Parthasarathy et al.\ 1983,
Cugier \& Hardrop 1988, Cugier 1989, Tomkin et al.\ 1993). This is in line with expectations
for the CNO cycle, which is dominant during the early evolution of a high-mass star
(Przybilla et al.\ 2010). However, for Algol systems, this picture may be altered
depending on the initial conditions, component masses and the mass ratio (Sarna 1992).
If a deep convective layer has developed, a standard cosmic abundance pattern is expected
instead (Sarna \& de Greve 1994).

\begin{figure*}
\includegraphics[width=18cm]{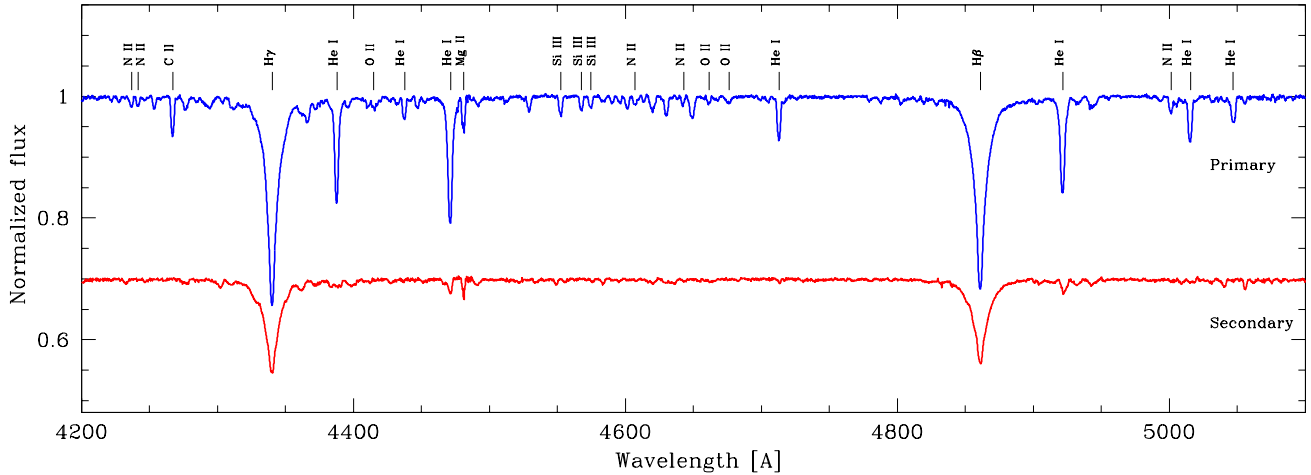}
\caption{ The portion of disentangled spectra of both components of the binary system
u Her in the region of the Balmer lines H$\gamma$ and H$\beta$. The spectrum of the
secondary component was shifted down by 0.3 for clarity. The lines of hydrogen, helium and 
some metals used in the analysis of the primary component are labelled.  }
\label{spdplot}
\end{figure*}

Early observational studies were hampered by line blending and/or the relative faintness
of the secondary star. Therefore a very limited line list was studied. The methods of
spectral disentangling ({\sc spd}; Simon \& Sturm 1994, Hadrava 1995) and Doppler
tomography (Bagnuolo \& Gies 1991), in conjunction with big advances in high-resolution
spectrographs, now make possible separation of the individual spectra of the components.
These disentangled spectra in turn make feasible a precise determination of the components'
effective temperatures (\Teff s) and photospheric chemical abundances, as elaborated by
Hensberge, Pavlovski \& Verschueren (2000) and Pavlovski \& Hensberge (2005).

As already stated, the photospheric abundance pattern in mass-transfer binary systems
preserves information on their past history. The initial characteristics of these systems
vary,  and a fine spectroscopic analysis of the abundance patterns can
provide additional evidence for proper discrimination
between different evolutionary paths and mass loss mechanisms (e.g.\ what fraction of
mass loss is by stellar wind, are mass loss and angular momentum changes conservative
or non-conservative, etc). With this aim in mind we initiated an observational project
of high-resolution \'echelle spectroscopy of bright Algol-type (semi-detached) binary
systems.

The binary system u\,Her (68\,Her, HD\,156633) belongs to a small group of early-type
semidetached systems first recognised by Eaton (1978). It differs from normal Algols
in several aspects: (i) the total mass is larger; (ii) the components are more similar
in \Teff; and (iii) the mass ratio is larger (Hilditch 1984). Their evolutionary paths
might also differ from those of normal Algols, which are the product of case B mass
transfer: it is supposed that in `hot Algols' case A mass transfer is involved (Webbink
1976). This was supported by the theoretical calculations of Nelson \& Eggleton (2001).

u\,Her is an eclipsing and double-lined spectroscopic binary with a rich observational
history thanks to its brightness ($V = 4.80$ mag at maximum light). The most recent
studies of u\,Her are those of Hilditch (2005) and Saad \& Nouh (2011). Both studies
contributed with new spectroscopic observations, but their measured stellar masses
differ. Whilst the masses for the components in Saad \& Nouh (2011) are similar to
earlier determinations (c.f.\ Kovachev \& Seggewiss 1975, Hilditch 1984) , Hilditch's
(2005) revised values are considerably greater, by $2.0 \pm 0.7$\Msun\ and $0.6 \pm
0.3$\Msun\ for the primary (mass-gaining) and secondary (mass-losing) components,
respectively.

The carbon abundance for the primary star has been estimated in two studies, which
disagree. First, Cugier (1989) analysed UV spectra obtained with the International
Ultraviolet Explorer (IUE) satellite for a group of Algols, and concluded that u\,Her
shows an essentially cosmic abundance of carbon. Contrarily, an analysis of optical
spectra by Tomkin, Lambert \& Lemke (1993) revealed a carbon deficiency in the primary
star with respect to the average carbon abundance of single B-type standard stars.

The primary goal of our study is the determination of the photospheric chemical
composition for the primary component in u\,Her. We have secured a new series of the
high-resolution \'echelle spectra and used {\sc spd} to isolate the spectra of the
two components. This enables us to determine the atmospheric parameters and the
elemental abundances from the entire optical spectral range. As a by-product the
two masses were also derived and compared to the previous solutions.
In Sect.\,\ref{sec:evo} an overview of the evolutionary calculations is presented,
and a possible evolutionary path for the components is discussed. The observed
[N/C] abundance ratio strengthens our conclusions from the model calculations.

\section{Spectroscopy}                                \label{sec:obs}

We obtained 43 spectra of u\,Her in the course of two observing runs (May and August 2008)
at the Centro Astron\'omico Hispano Alem\'an (CAHA) at Calar Alto, Spain. We used the
2.2\,m telescope, FOCES \'echelle spectrograph (Pfeiffer et al.\ 1998), and a Loral
\#11i CCD binned 2$\times$2 to decrease the readout time. With a grating angle of 2724,
prism angle of 130 and a 150\,$\mu$m slit we obtained a spectral coverage of roughly
3700--9200\,\AA\ in each exposure, at a resolving power of $R \approx 40\,000$.
Wavelength calibration was performed using thorium-argon exposures, and flat-fields
were obtained using a tungsten lamp. The observing conditions were generally good
but several exposures suffered from the presence of thin clouds.

The \'echelle spectra were bias-subtracted, flat-fielded and extracted with the
{\sc iraf}\footnote{{\sc iraf} is distributed by the National Optical Astronomy
Observatory, which are operated by the Association of the Universities for Research
in Astronomy, Inc., under cooperative agreement with the NSF.} \'echelle package
routines. Normalisation and merging of the orders was performed with great care,
using custom programs, to ensure that these steps did not cause any systematic
errors in the resulting spectra.

\begin{figure}
\includegraphics[width=9cm]{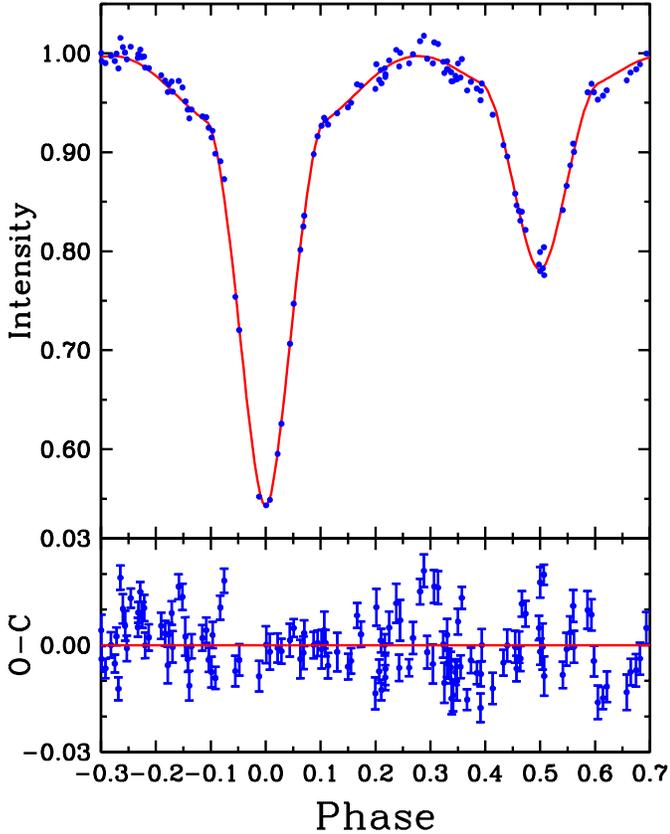}
\caption{Observed phased Hipparcos $H_{\rm p}$-band light-curves of u\,Her with the
best-fitting {\sc phoebe}
model light curves. The error bars are of similar size to the data points. In the
lower panel the residuals have been plotted to show the goodness of the fit.}
\label{figLC}
\end{figure}

\section{Spectroscopic orbit through spectral disentangling}            \label{sec:orbits}

In {\sc spd} the individual component spectra are isolated simultaneously with the
determination of the optimal set of orbital elements. The reliability of the separated
spectra and orbital elements depends on the spectral characteristics of the components
and their contribution to the total light of the system.

According to previous photometric solutions for u\,Her (c.f.\ Hilditch 2005), the
primary star is about 3.3 times brighter than the secondary. Also, their \Teff s are
quite different. For the primary star ($\Teff \sim 22\,000$\,K) the \ion{He}{i} lines
are expected to be at maximum strength. For the secondary star ($\Teff \sim 12\,500$\,K)
the \ion{He}{i} lines should be quite weak but the Balmer lines stronger than for the
primary. Metal lines, quite prominent in the primary, are much weaker in the secondary,
both intrinsically and due to the component's relative faintness. Therefore, we concentrate
on the Balmer and helium lines, putting more emphasis on the latter.

The Fourier method (Hadrava 1995), is the best choice for disentangling spectra covering
an extensive spectral range with high spectral resolution. The code
{\sc FDBinary}\footnote{Available at {\tt  http://sail.zpf.fer.hr/fdbinary/}}
(Iliji\'{c} et al.\ 2004) was used to perform {\sc spd} in spectral regions centred
on the prominent helium and Balmer lines, covering 50--150\,\AA\ in each region (Fig.~1).
The orbital solution obtained by {\sc spd} yields velocity amplitudes of $K_1 = 94.6
\pm 2.3$\kms\ and $K_2 = 267.4\pm3.3$\kms, and thus a mass ratio of $q = 0.354\pm0.010$,
under the assumption of a circular orbit.

In both recent studies of u\,Her a spectroscopic orbit was determined by {\sc spd},
but the results are in astonishingly poor agreement for such a bright object. Our values
for the two velocity amplitudes are much closer to those found by Saad \& Nouh (2011)
and in clear disagreement with those from Hilditch (2005).  Saad \& Nouh (2011) based
their solution on red-optical spectra covering H$\alpha$ and the \ion{He}{i} 6678\,\AA\
line, finding $K_1 = 98$\kms\ and $K_2 = 265$\kms\ (no errors are quoted). Hilditch
(2005) used grating spectra covering 450\,\AA\ of the blue-optical spectral region,
finding $K_1 = 102.4\pm2.4$\kms\ and $K_2 = 274.8\pm0.9$\kms. In particular $K_2$
is considerably higher than earlier measurements. We suspect that the disagreement
between the orbital solutions stems primarily from the different spectral resolution
employed; Hilditch's grating spectra have a resolution of 0.46\,\AA/px, while our
\'echelle spectra have a much higher resolution of 0.02\,\AA/px. However, the study
by Kovachev \& Seggewiss (1975) yielded $K_1 = 95.6\pm1.4$\kms\ and $K_2 = 263\pm3$\kms,
from photographic spectra of a similar resolution to Hilditch's digital spectra, and
RV measurements by the classical oscilloscopic method.

Discrepancies in the masses calculated from the above spectroscopic solutions are
more pronounced for the primary, $M_1 = 7.9$ to $8.8$\Msun, than for the secondary,
$M_2 = 2.8$ to $3.3$\Msun. In his final solution Hilditch (2005) corrected the two
velocity amplitudes for the distorted shape of the stars and their mutual irradiation,
resulting in masses of $M_1 = 9.61\pm0.14$\Msun\ and $M_2 = 3.48\pm0.13$\Msun.
These are, as noted by Hilditch, considerably higher than in earlier solutions for u\,Her.

\section{Light curve modelling}

Since u\,Her is a bright object it has a long history of photometric measurements.
All published ground-based light curves show night-to-night variations and a scatter
of about 0.04\,mag (S\"{o}derhjelm 1978, Rovithis \& Rovithis-Livaniou 1980, van der
Veen 1984). In contrast, the {\it Hipparcos} satellite photometry ($H_p$ passband)
covers about 1160 days between 1989 and 1993, and is of good quality
(Fig.\,2). Hilditch (2005) discussed possible explanations for the erratic
night-to-night variations and concluded that they are intrinsic to the system. Since
no periodicity could be determined he flagged it as a semi-regular variable. He asserted
that the time coverage of the {\it Hipparcos} photometry corresponds to a quiescent
period of the system. Therefore, we decided to reanalyse only the $H_p$ photometry.

In a period analysis of ten semi-detached Algol-type binaries, \.{I}bano\v{g}lu et
al.\ (2012) found u\,Her
 to be the system with the smallest period changes.  We used
their ephemeris\footnote{In Table 6 of  \.{I}bano\v{g}lu et al.\ (2012) there is
a typo, confusing u\,Her with U\,Her, a common mistake in the literature, see Hilditch (2005).}
 during our analysis; $T_{\rm prim. min.}({\rm HJD}) = 2447611.5007(15) +
2.05102685(68)\times E$ where the standard deviations in the last significant digits
are given in brackets.

Initially we set the primary's \Teff\ to $20\,000$\,K, as derived by Hilditch (2005)
using $uvby$ photometry and the $[u\!-\!b]-T_{\rm eff}$ calibration from Tomkin et al.\
(1993) and Napiwotzki et al.\ (1993). In a second iteration we fixed it at our revised
value of $\Teff = 21\,600\pm220$ K (Sect.\,5.1). Our value for the primary  \Teff\
agrees within $1\sigma$ with Hilditch's, who also noted that there is no significant
effect on the light curve solutions when values between 19\,000 and 21\,500\,K are used.

\begin{figure}
\includegraphics[width=9cm]{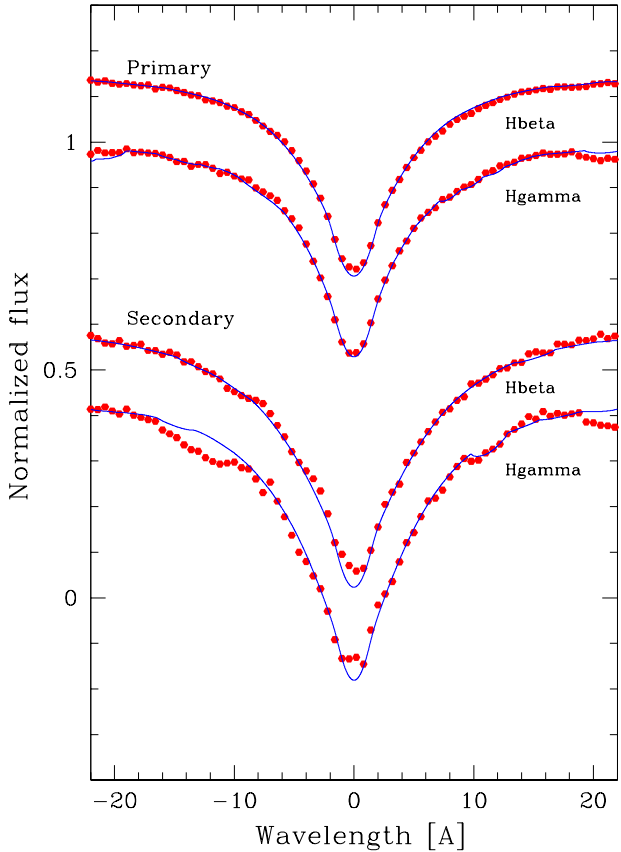}
\caption{\label{fig:betagamma} The best-fitting synthetic spectra (lines) compared
to the renormalised disentangled spectra (filled circles) of the two stars. For
both components H$\beta$ (upper) and H$\gamma$ (lower) profiles are shown.}
\end{figure}

In order to analyse the light curve we deployed the Wilson-Devinney ({\sc wd}) code
implemented into the {\sc phoebe} package by Pr{\v s}a \& Zwitter (2005). Our initial
unconstrained system parameters immediately converged to a semi-detached configuration.
Thus, we used the {\sc mode=5} option for subsequent solutions. Since {\sc wd} uses
Roche geometry, the solutions are sensitive to the mass ratio, which we fixed at the
value derived from our orbital solution above, $q = 0.354 \pm 0.010$.
Other fixed parameters in our light curve calculations are listed in
Table\,1.

The orbital inclination, secondary-star \Teff, primary-star surface potential, phase shift,
and fractional primary-star luminosity were put as adjustable parameters. We kept the other
parameters as fixed. Iterations were carried out until convergence was achieved. The formal
 parameter uncertainties calculated by {\sc wd}'s differential correction solver ({\sc dc})
are not trustworthy, so we implemented a more robust approach to error estimation. We
simulated a range of solution sets around our fixed parameter values and calculated the
$\chi^2$ value of each. Then we accepted each parameter's $2\sigma$ confidence level as
our error range, assuming that the $\chi^2$ values follow a Gaussian distribution.
In Table\,1 we show the final parameter set and their corresponding
error estimations. The computed light curve and residuals from observations is
shown in Fig.\,2.

\begin{table} \centering
\caption{Results from the solution of $H_p$ band light-curves of u\,Her.}
\label{parameters}
\begin{tabular}{llr}
\hline
Parameter                  & Unit & Value                   \\
\hline
{\em Fixed parameters:}    &      &                         \\
Orbital period $P$         & d    & $2.05102685$            \\
Primary eclipse time HJD   & d    & $2\,447\,611.5007$      \\
Mass ratio $q$             &      & $0.354$                 \\
$T_{\rm eff}$ of star A    & K    & $21\,600$               \\
Primary LD coefficients    &      & 0.434,0.252             \\
Secondary LD coefficient   &      & 0.568,0.318             \\
Gravity darkening          &      & $1.0$, $1.0$            \\
Bolometric albedo          &      & $1.0$, $1.0$            \\
Third light                &      & $0.0$                   \\[3pt]
{\em Fitted parameters:}   &      &                         \\
Star A potential           &      & $3.437 \pm 0.250$       \\
Orbital inclination        & deg  & $78.9 \pm 0.4$           \\[3pt]
$T_{\rm eff}$ of star B    & K    & $12\,700 \pm 140$       \\
{\em Derived parameters:}  &      &                         \\
$L_{1}/(L_1+L_2) $         &      & $0.739 \pm 0.026$       \\
Fractional radius of star A &     & $0.330 \pm 0.009$       \\
Fractional radius of star B &     & $0.285$ (fixed)         \\
\hline
\end{tabular}
\end{table}

\begin{figure}
\begin{tabular}{c}
\includegraphics[width=85mm]{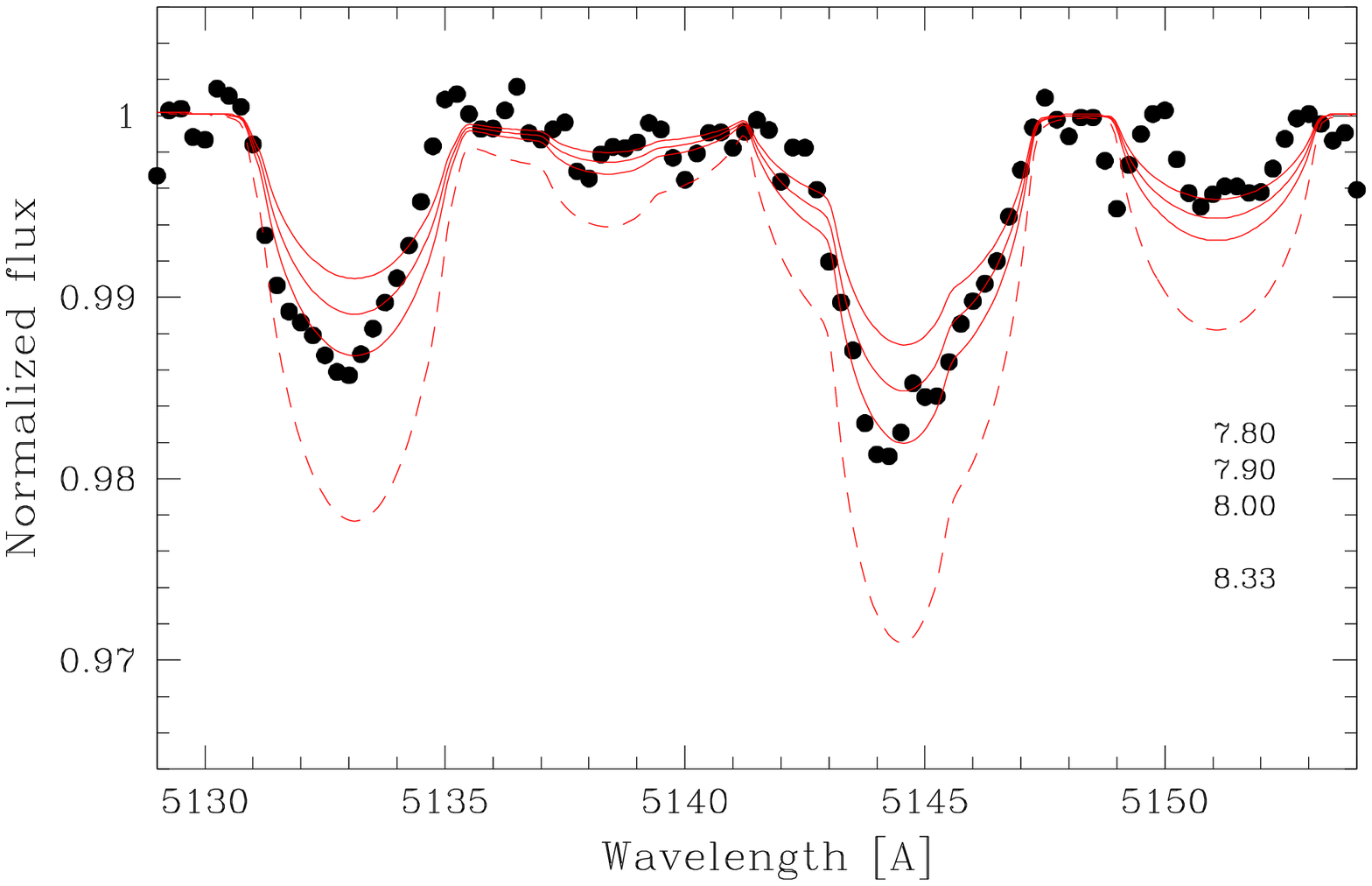} \\
\includegraphics[width=85mm]{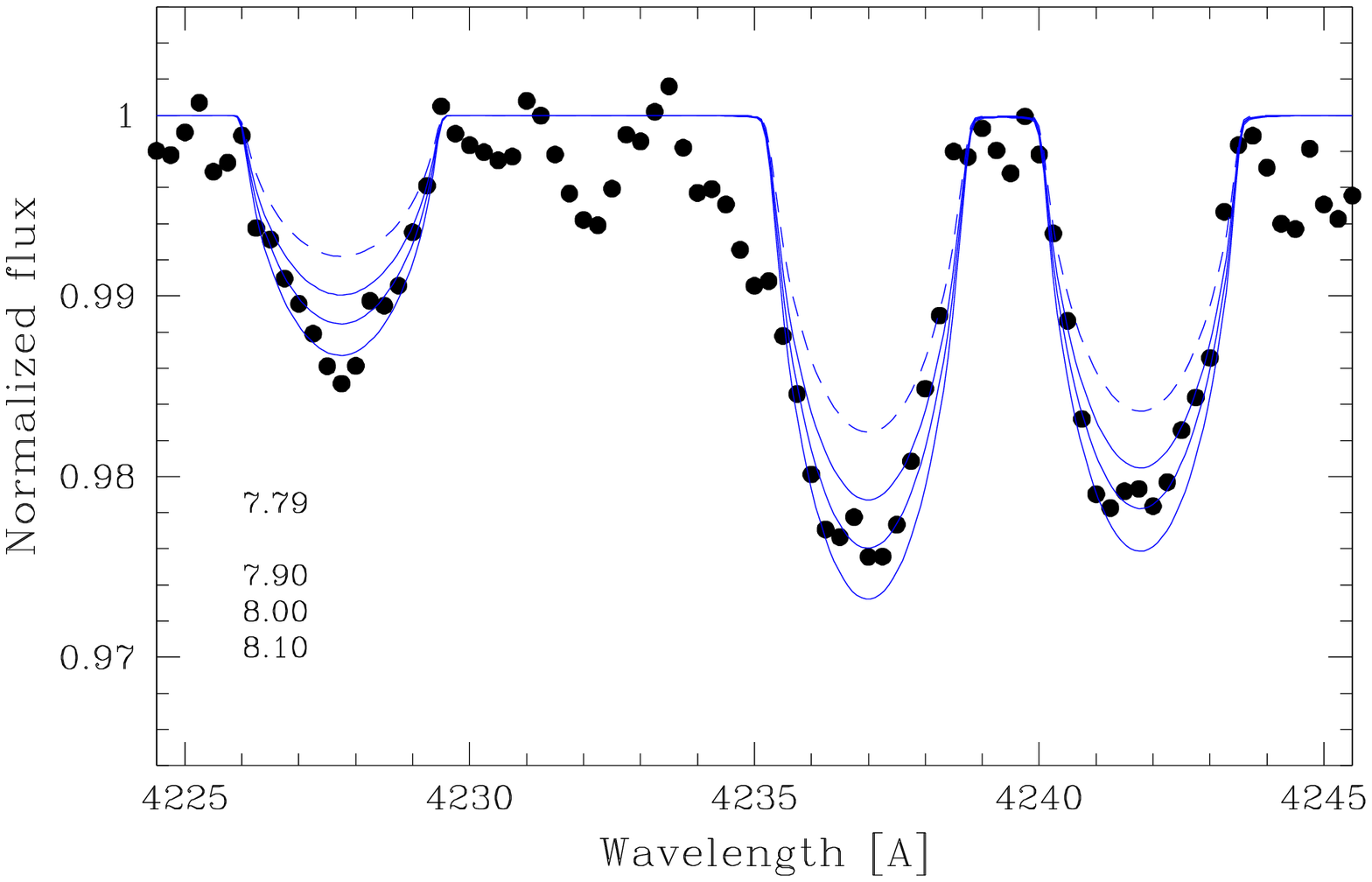} \\
\includegraphics[width=85mm]{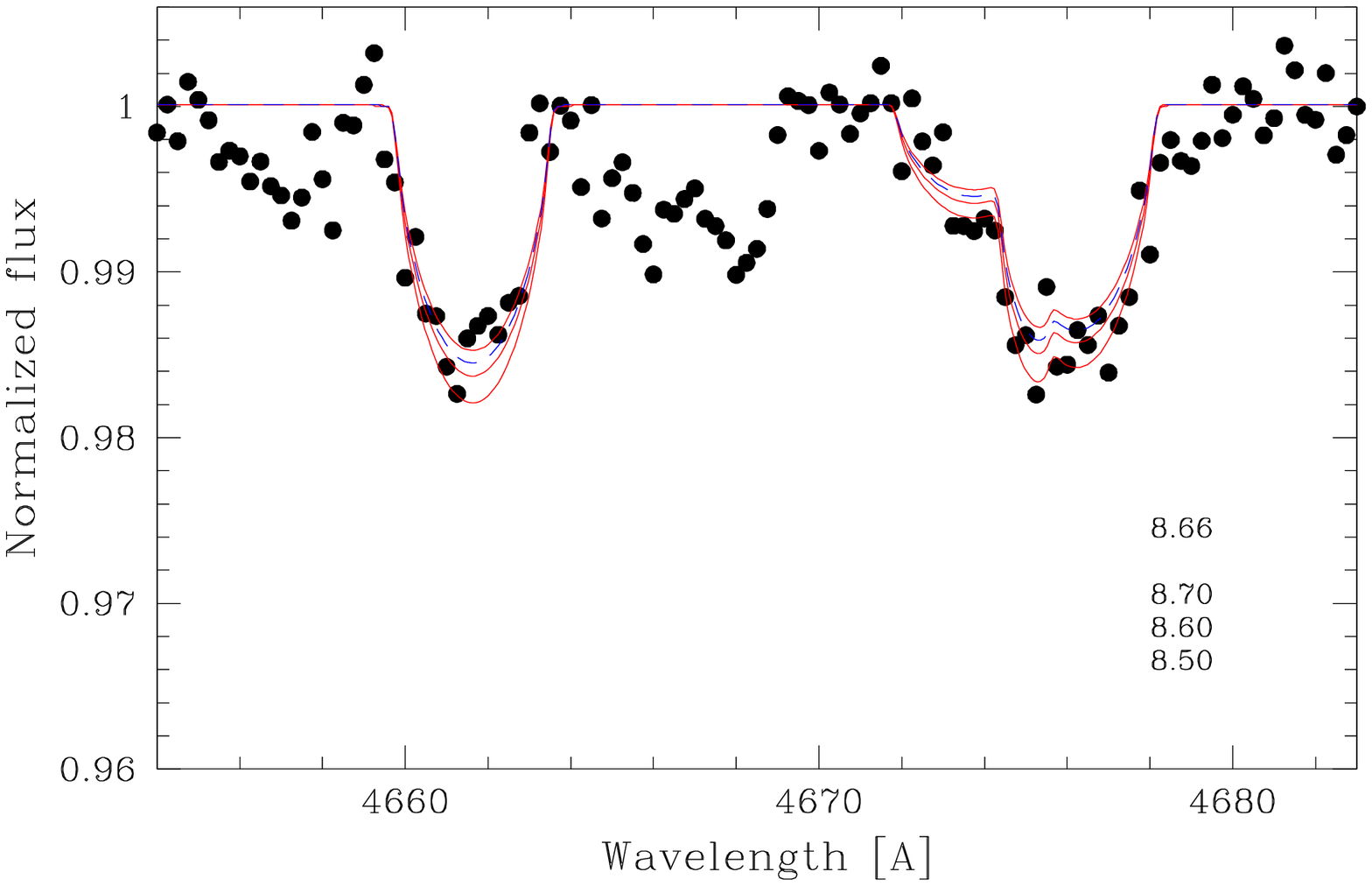} \\
\end{tabular}
\caption{\label{fig:plotspe} Comparison between the renormalised disentangled spectrum
of the primary component in u\,Her (filled circles) and a grid of theoretical spectra
computed assuming different abundances. From top to bottom, the panels show
selected lines for carbon, nitrogen, and oxygen, respectively. The abundances used for 
the theoretical spectra
are indicated in the labels. Spectra calculated for the `present-day cosmic
abundances' for the Galactic OB stars (Nieva \& Przybilla 2012) are indicated
by dashed lines.}
\end{figure}

\begin{figure}
\begin{tabular}{cc}
\includegraphics[width=38mm]{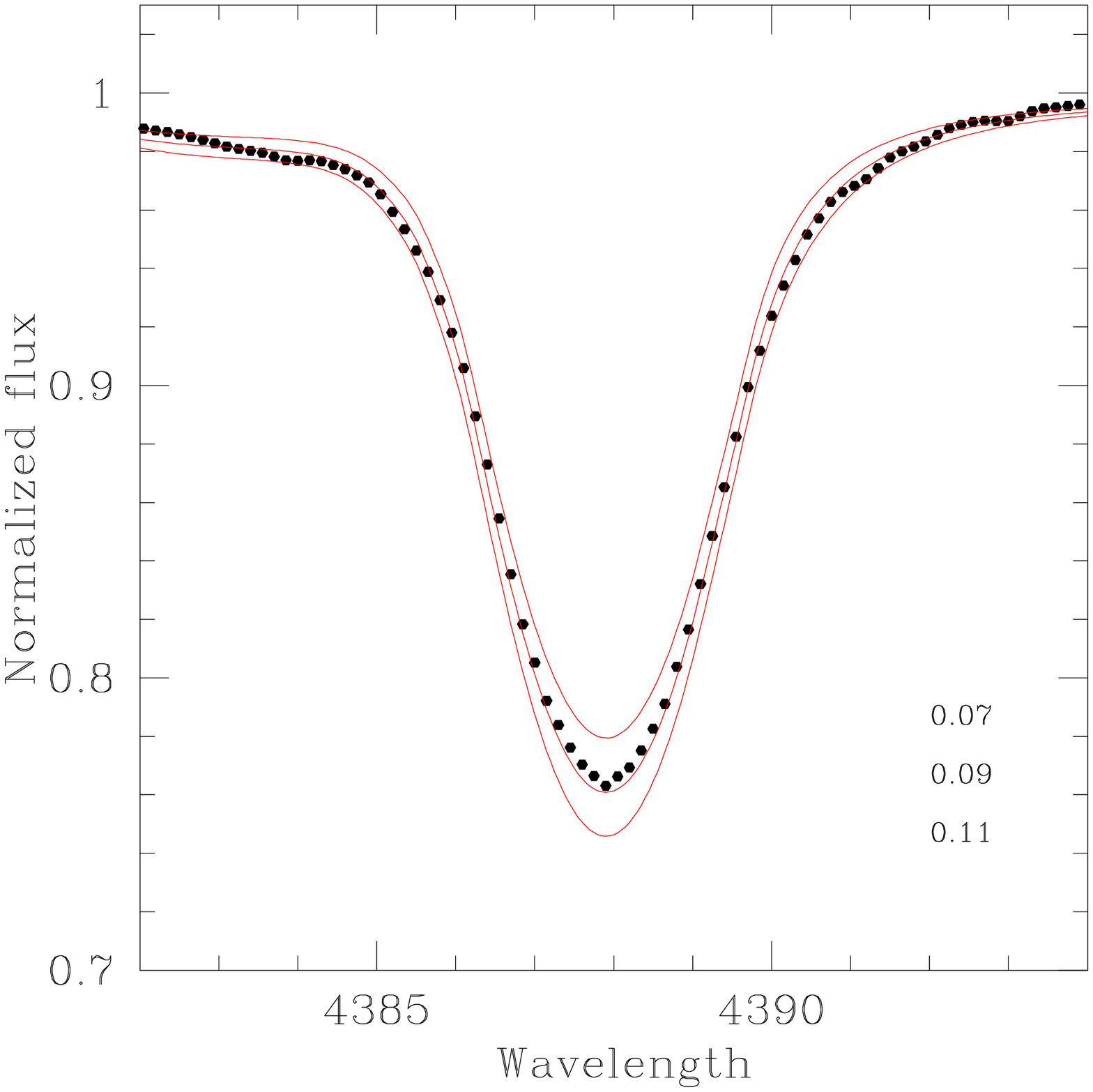} &
\includegraphics[width=38mm]{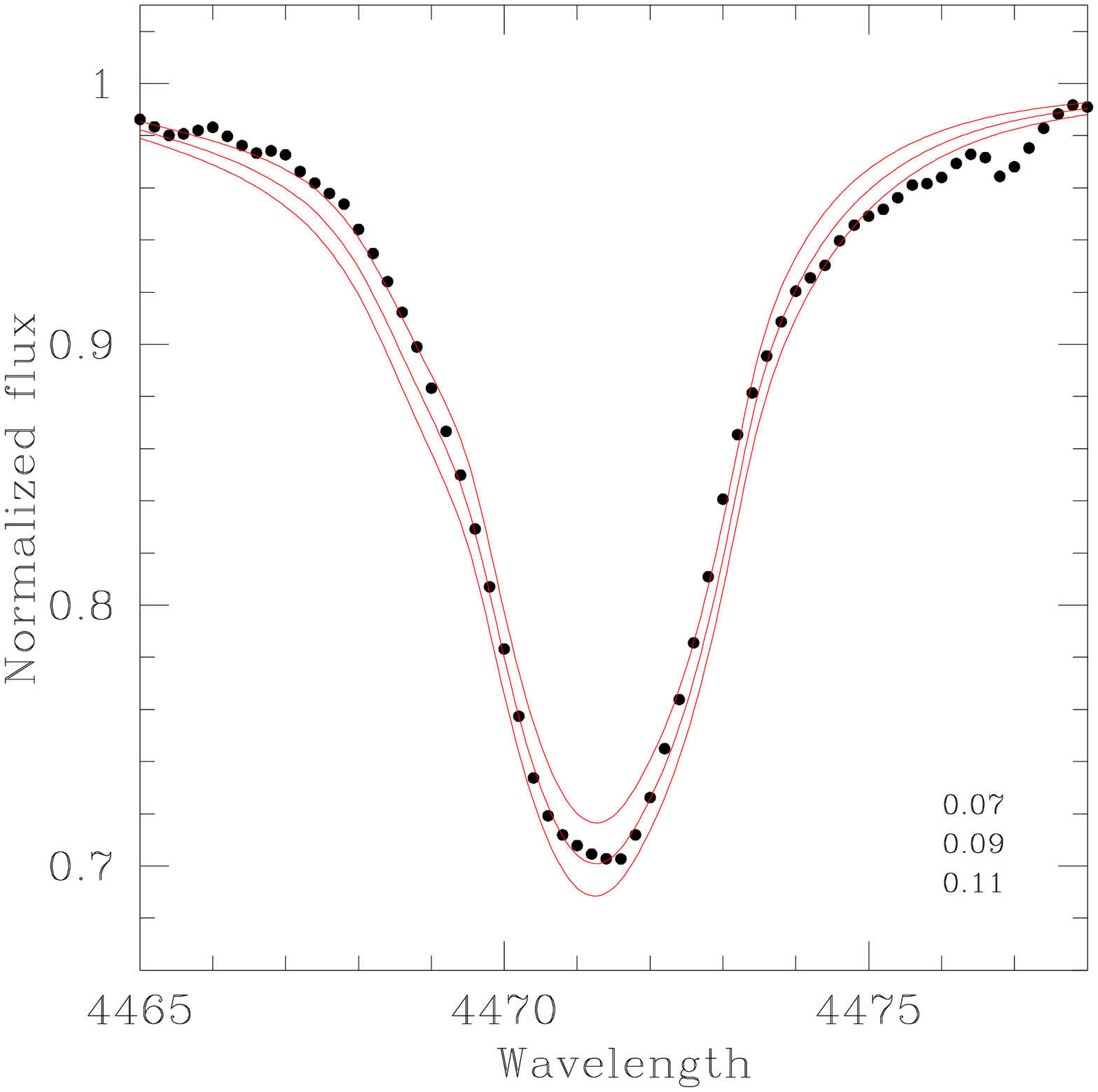} \\
\includegraphics[width=38mm]{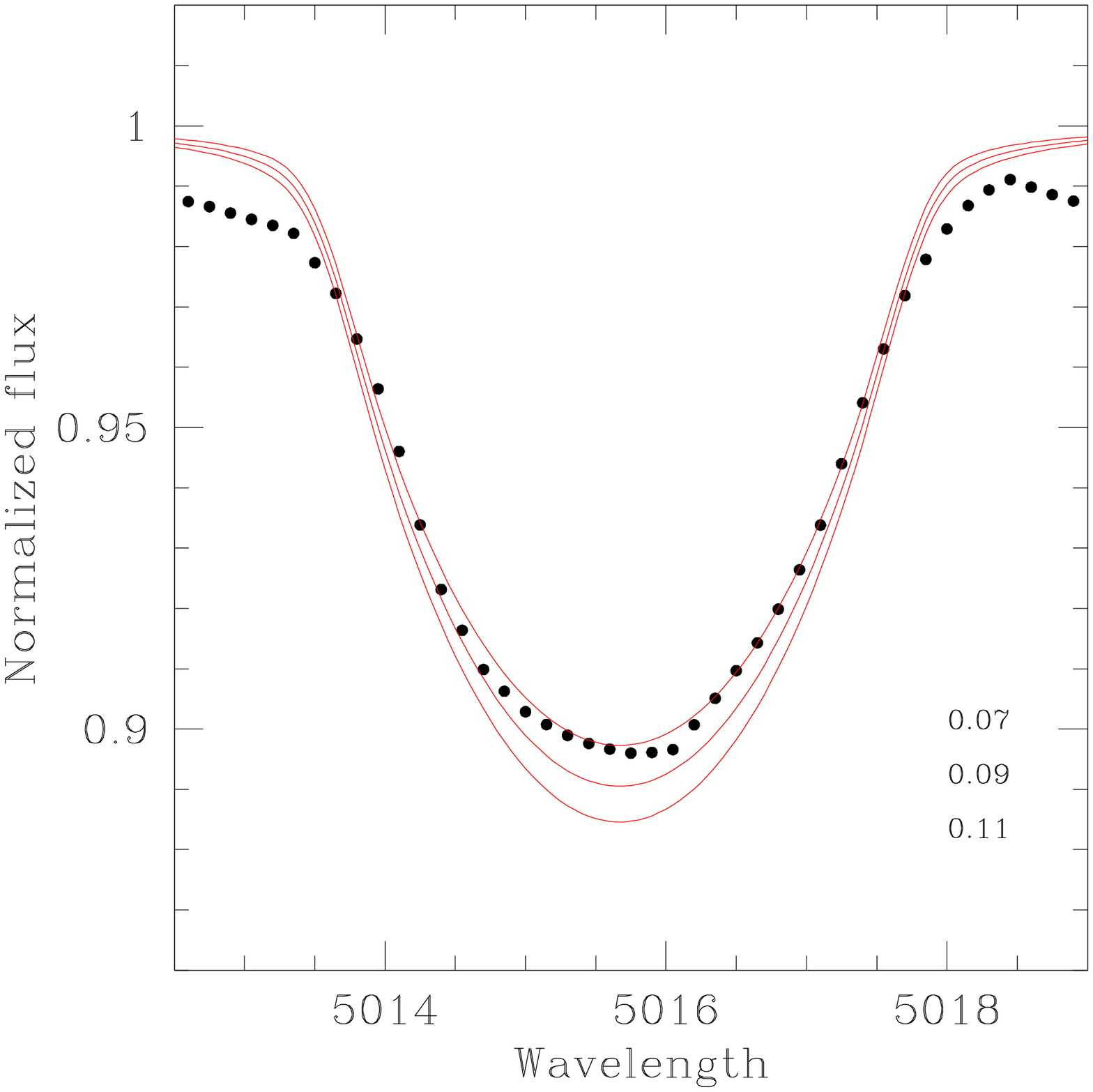} &
\includegraphics[width=38mm]{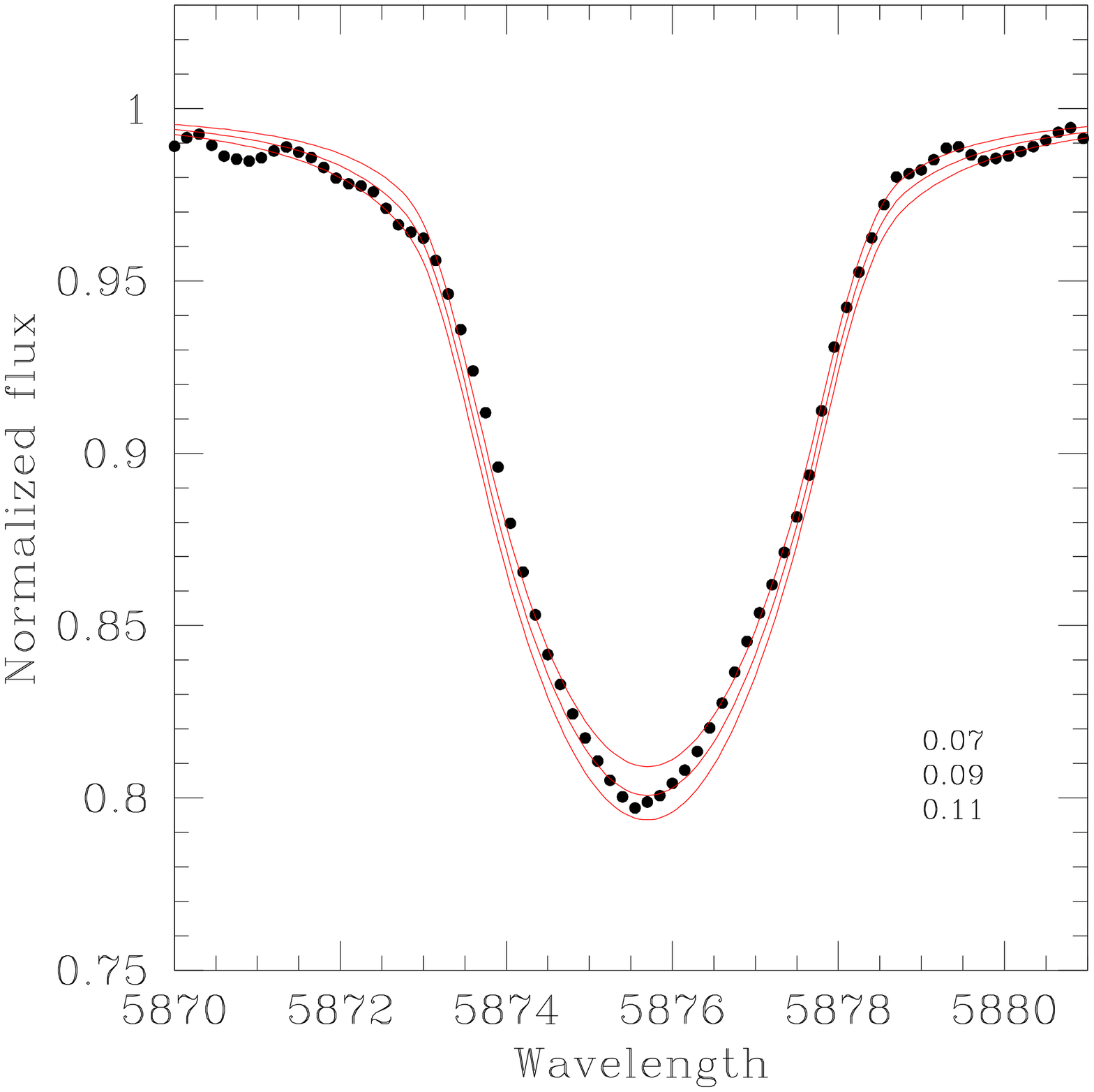} \\
\end{tabular}
\caption{
 The quality of the fits for selected \ion{He}{i} lines at 4387.9, 4471.5, 5015.7 and 5876.7\,\AA\ 
to the theoretical spectra
(lines). Theoretical spectra were calculated for the stellar parameters listed in Table\,2;
the assumed helium abundances are indicated in the bottom-right corner of each panel, and are
expressed by a fraction of the helium atoms to the total number of atoms
in the stellar atmosphere. The renormalised disentangled spectra are represented
by filled circles.}
\label{helplot}
\end{figure}

\section{Spectral analysis of both components}

\subsection{The effective temperatures}

To construct model atmospheres for the individual components of binary system, we first
need to set their \Teff s and surface gravities (\logg s). When stars are in binary
systems for which accurate masses and radii can be derived from radial velocity and
light curves, the resulting \logg\ measurements have a much higher precision that
those determinable from spectroscopic analysis alone. In the case of u\,Her, and even
though our observational data give only a modest accuracy in the masses and radii
(about 3--4\% and 2--3\% respectively) we measured \logg\ values to about 0.013\,dex
for the primary and and 0.018\,dex for the secondary.

The availability of these \logg\ measurements lifts the degeneracy between \Teff\ and
\logg\ as determined from Balmer line profiles. However, the trade-off is that the
disentangled spectra of the components must be renormalised to their intrinsic continuum
flux. Pure {\sc spd} returns the components' spectra relative to a common continuum level,
 and the individual spectra of the components are diluted by the factor proportional to
their fractional contribution to the total light of the system. If no input spectra were
obtained during eclipse (i.e.\ the component light ratio is the same for all spectra),
the zeroth-order mode in the Fourier disentangling is singular, and an ambiguity in the
proper renormalisation of the disentangled spectra to their own continuum appears
(Pavlovski \& Hensberge 2005). Hence, external information on the light ratio is needed
(see Pavlovski \& Hensberge 2010 and Pavlovski \& Southworth 2012). In the case of u\,Her
 we can use the light ratio as determined by the light curve solution (Sect.\,4),
${l_1}/{l_2} = 0.300 \pm 0.003$, where $l_1$ and $l_2$ are the fractional contributions
of the components to the total light of the system.

The optimal fitting of the Balmer lines in the renormalised component spectra to the grid
of synthetic spectra was performed with our {\sc starfit} code. This optimisation routine
uses a genetic algorithm inspired by the {\sc pikaia} subroutine of Charbonneau (1995).
The following parameters for each component can be either optimised or fixed: \Teff,
\logg, light factor, projected rotational velocity (\Vsini), velocity shift, and
continuum level adjustment. The velocity shift is needed because in {\spd} there is
no absolute wavelength scale, and disentangled spectra are returned on a wavelength
scale with an arbitrary zero point. The reason for the continuum level adjustment is
the fact that disentangled spectra are shifted according to the line blocking of the
individual components, and an additive constant is needed to rectify disentangled
spectra to a continuum of unity (Pavlovski \& Hensberge 2005). This is the main improvement
over our previous code {\sc genfit} (Tamajo et al.\ 2011). {\sc starfit} can be
run in constrained mode (simultaneous fit for both components with the condition that
$l_1 + l_2 = 1.0$) or unconstrained mode (see Tamajo et al.\ 2011). For u\,Her we ran
{\sc starfit} in unconstrained mode with the light ratio fixed to that from the light
curve model and the \logg\ values of the stars fixed. Since the intrinsically broad Balmer
lines are almost unaffected by the rotational kernel, the \Vsini\ values for both
components were first derived by iteratively fitting helium and metal lines. The
$\Teff$s were then determined from the optimal fitting of the H$\gamma$ and H$\beta$
lines. The results for $\Teff$ and \Vsini\ for both components are given in
Table\,3. The quality of the fits is illustrated in
Fig.\,3. The spectroscopically-determined \Teff\ for the secondary
component is in perfect agreement with the results from the
light curve analysis (c.f.~Table\,1).

\begin{table} \centering \caption{\label{tab:dimensions} The
absolute dimensions and related quantities determined for u\,Her.
\Vsync\ is the calculated synchronous rotational velocity.}
\begin{tabular}{l c r@{\,$\pm$\,}l r@{\,$\pm$\,}l} \hline
 Parameter     & Unit      &      \mc{Star A}    &      \mc{Star B}    \\
\hline
Semimajor axis & \Rsun     & \multicolumn{4}{c}{$14.95 \pm 0.17$}      \\
Mass           & \Msun     & $7.88$    & $0.26$  & $2.79$    & $0.12$  \\
Radius         & \Rsun     & $4.93$    & $0.15$  & $4.26$    & $0.06$  \\
\logg\         & \cmss     & $3.948$   & $0.024$ & $3.625$   & $0.013$ \\
\Teff          & K         & $21\,600$ & $220$   & $12\,600$ & $550$   \\
$\log L$       & L$_\odot$ & $3.68$    & $0.03$  & $2.63$    & $0.08$  \\
\Veq\ $\sin i$ & \kms      & $124.2$   & $1.8$   & $107.0$   & $2.0$   \\
\Vsync\        & \kms      & $121.7$   & $3.5$   & $105.0$   & $1.5$   \\
\hline \end{tabular} \end{table}

\begin{figure}
\includegraphics[width=9cm]{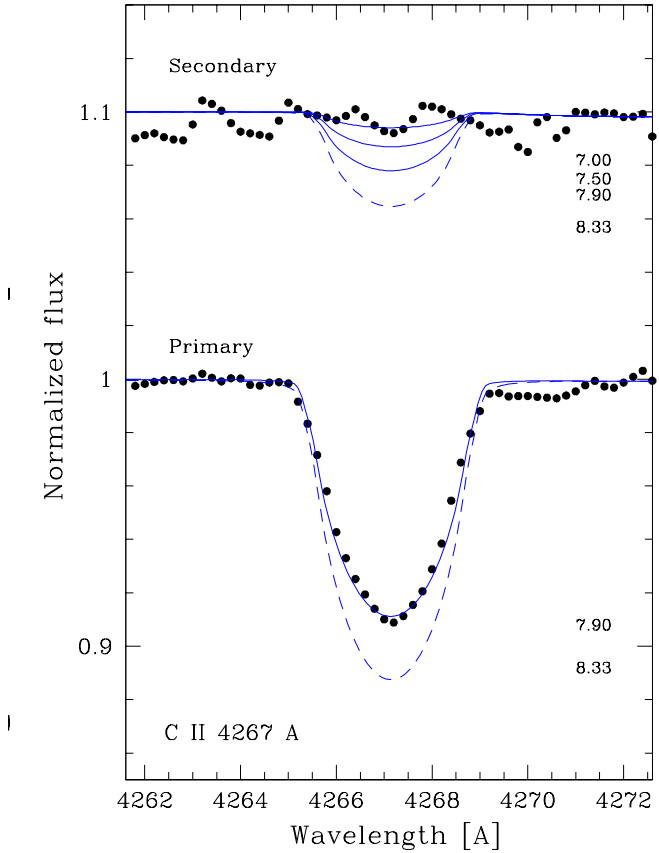}
\caption{\label{fig:carbon}
 Renormalised line profiles of \ion{C}{ii} 4267\,\AA\ for the components of
u\,Her. The secondary spectrum is shifted by +0.1
for clarity. Theoretical spectra calculated for
different abundances (indicated in the labels) are shown for comparison.
Dashed lines represent the `present-day cosmic abundance' of carbon
derived by Nieva \& Przybilla (2012). Whilst the carbon abundance is
depleted in the primary, it is almost an order of magnitude less
in the atmosphere of the secondary component. See Sect.\,5.2 for details.}
\label{carbon}
\end{figure}

\subsection{Abundances}

Theoretical spectra for the atmospheric parameters of the primary and varying
microturbulence velocities (\micro) and elemental abundances were calculated
in a `hybrid' approach (Nieva \& Przybilla 2007, Przybilla et al.\ 2010), which
combines local thermodynamic equilibrium (LTE) atmospheres and non-LTE line-formation
calculations. We computed model atmospheres with the {\sc atlas9} code, which assumes
plane-parallel geometry, chemical homogeneity, and hydrostatic, radiative and local
thermodynamic equilibrium. Line blanketing was realised by means of opacity distribution
functions (ODFs). Solar abundances were adopted in all computations. Non-LTE level
populations and model spectra were obtained with the {\sc detail} and {\sc surface}
codes (Giddings 1981, Butler \& Giddings 1985). Non-LTE level populations and synthetic
spectra of H, He, C, N, O, Mg, Si and Al were computed using the most recent model atoms
(see Table\,3 in Nieva \& Przybilla 2012). { The \micro\
was determined from the condition of null-correlation between O abundance and
equivalent width. Oxygen is used for this purpose since the O lines are the
most numerous in spectra for \Teff s similar to that of the primary component.
Only the lines selected by Sim\'{o}n-D\'{i}az (2010) were used. The \micro\
determined, $2 \pm 1$\kms, is in the range of typical values for early-B
type stars on the main sequence (c.f.\ Simon-Diaz 2010, Nieva 2011, Nieva \&
Sim\'{o}n-D\'{i}az 2011, Nieva \& Przybilla 2012). Abundances are estimated by
line profile fitting and are listed for all transitions calculated and for lines which are not
severely blended. They are given in Table\,3 for \ion{He}{i} and Table\,4 for
all other ions. The mean values of abundances and their
uncertainties for all elements studied in the disentangled spectrum of the primary
component are given in Table\,5.
The uncertainties in the
abundances are calculated from the scatter in the estimated abundances for
different lines, and for 1$\sigma$ deviations in the \Teff\
and \micro. However, the prevailing uncertainty in abundances
comes from the scatter between different lines.

\begin{table} \centering \caption{\label{tab:hellines}
 Photospheric helium abundance for the primary component of u Her
as derived from different \ion{He}{i} spectral lines.
The abundances are given as the fractional number of helium
atoms to the total number of atoms  in the stellar
atmosphere  $(N({\rm H})+N({\rm He}))$   .}
\begin{tabular}{lclc}
\hline
Line  & $N$(He)  & Line & $N$(He)         \\
\hline
4387.9 & 0.087 $\pm$ 0.008 & 5015.7 & 0.075 $\pm$ 0.006  \\
4437.6 & 0.101 $\pm$ 0.010 & 5047.7 & 0.091 $\pm$ 0.007  \\
4471.5 & 0.088 $\pm$ 0.007 & 5876.7 & 0.095 $\pm$ 0.002  \\
4713.2 & 0.078 $\pm$ 0.006 & 6678.1 & 0.086 $\pm$ 0.007  \\
4921.9 & 0.104 $\pm$ 0.005 &             &        \\
\hline
\end{tabular}
\end{table}

\begin{table} \centering \caption{\label{tab:lines}
 Estimated photospheric abundances for different ions in the
atmosphere of the primary component of u\,Her.
Abundances are expressed relative to the abundance of hydrogen,
$\log \epsilon({\rm H}) = 12.0$.}
\begin{tabular}{lclclc}
\hline
Line  & $\log \epsilon({\rm X})$ & Line & $\log \epsilon({\rm X})$ & Line  & $\log \epsilon({\rm X})$  \\
\hline
\ion{C}{ii}& & \ion{O}{ii} &  & \ion{Mg}{ii} &     \\
4267.00 & 7.90 & 4185.46 & 8.55 & 4481.13  & 7.50 \\
5132.95 & 7.95 & 4189.79 & 8.70 & 5264.22  & 7.40 \\
5137.26 & 7.97 & 4414.88 & 8.55 & 5401.54  & 7.50 \\
5143.40 & 7.95 & 4416.97 & 8.62 & \ion{Si}{ii} &  \\
5151.08 & 7.85 & 4590.97 & 8.70 & 4128.05  & 7.55   \\
\ion{N}{ii}&   & 4596.20 & 8.68 & 4130.88  & 7.40   \\
4227.75 & 8.00 & 4609.42 & 8.50 & \ion{Si}{iii} & \\
4236.91 & 7.98 & 4661.63 & 8.62 & 4552.62  & 7.48  \\
4241.80 & 7.95 & 4673.75 & 8.65 & 4567.82  & 7.37  \\
4447.03 & 8.00 & 4676.23 & 8.65 & 4574.76  & 7.55  \\
4507.56 & 8.00 & 4677.07 & 8.70 & 4716.65  & 7.70  \\
4607.15 & 8.05 & 4698.48 & 8.60 & 4819.72  & 7.50  \\
4613.86 & 8.00 & 4699.21 & 8.68 & 4828.96  & 7.35  \\
4643.09 & 7.90 & 4703.18 & 8.48 & \ion{Al}{iii} & \\
4803.27 & 8.05 & 4705.35 & 8.55 & 4149.92  & 6.30  \\
4987.38 & 8.00 & 5206.64 & 8.50 & 4479.97  & 6.40  \\
4994.36 & 8.02 &         &      & 4512.54  & 6.20  \\
5001.47 & 7.95 &         &      & 4528.94  & 6.40  \\
5007.31 & 7.83 &         &      &          &       \\
5010.62 & 7.90 &         &      &          &       \\
5045.09 & 7.93 &         &      &          &       \\
5495.65 & 8.03 &         &      &          &       \\
5666.63 & 7.85 &         &      &          &       \\
\hline
\end{tabular}
\end{table}

\begin{table} \centering \caption{\label{tab:abuall} Photospheric abundances
derived for the primary component of u\,Her. Abundances are expressed relative
to the abundance of hydrogen, $\log \epsilon(H) = 12.0$. The third column gives
the number of lines used. Present-day cosmic abundances from Galactic OB stars
(Nieva \& Przybilla 2012) are given in the fourth column, and the fifth column
lists the solar abundances from Asplund et al.\ (2009).}
\begin{tabular}{lcrcccc} \hline
El.\  & $\log \epsilon({\rm X})$ & $N$ & [X/H] & OB stars      & Sun         \\
\hline
He & 10.99$\pm$0.05 & 6  &  0.02$\pm$0.05  & 10.99$\pm$0.01 & 10.97$\pm$0.01\\
C  & 7.92$\pm$0.02  & 5  & -0.47$\pm$0.05  & 8.33$\pm$0.04 & 8.39$\pm$0.05 \\
N  & 7.97$\pm$0.02  & 17 &  0.20$\pm$0.06  & 7.79$\pm$0.04 & 7.78$\pm$0.06 \\
O  & 8.61$\pm$0.02  & 16 & -0.05$\pm$0.05  & 8.76$\pm$0.05 & 8.66$\pm$0.05 \\
Mg & 7.47$\pm$0.03  & 3  & -0.06$\pm$0.09  & 7.56$\pm$0.05 & 7.53$\pm$0.09 \\
Si & 7.49$\pm$0.04  & 8  & -0.02$\pm$0.06  & 7.50$\pm$0.05 & 7.51$\pm$0.04 \\
Al & 6.32$\pm$0.05  & 4  & -0.05$\pm$0.06  &               & 6.37$\pm$0.04 \\
\hline \end{tabular} \end{table}

Cugier (1989) determined the carbon abundances in the mass-gaining components of six
Algol-type systems, including u\,Her. He used UV spectra taken with the IUE, and
measured the total equivalent widths of the \ion{C}{ii} multiplets at 1334.5--1335.7\,\AA\
and 1323.8--1324.0\,\AA. He constrained the components' \Teff s from the UV flux
distribution and van der Veen's (1983) photometric solution, finding $T_{\rm eff,1} =
22\,200\pm1500$\,K and $T_{\rm eff,2} = 13\,300\pm1000$\,K. After correction for
non-LTE effects he found $\log \epsilon {\rm(C)} = 8.62 \pm 0.30$, and concluded that
the primary of u\,Her shows no indication of a change in the carbon abundance, in contrast
 to other Algols in his sample.

Tomkin et al.\ (1993) reported high-resolution CCD spectra of the \ion{C}{ii} 4267\,\AA\
line in the same Algol systems that were studied by Cugier (1989) and Cugier \& Hardorp
(1988). They estimated carbon abundances in the Algol primaries differentially with
respect to single B-type stars. They derived the \Teff s from Str\"{o}mgren photometry
using the calibration by Napiwotski et al.\ (1993), finding $T_{\rm 1, eff} = 20\,000$\,K.
Tomkin et al.\ obtained a carbon abundance of [C/H] $= -0.34$ with respect to the average
abundance of the standard stars, $\log \epsilon {\rm(C)} = 8.28$.

The carbon abundance determined in this work, $\log \epsilon ({\rm C}) = 7.92\pm0.02$,
is based on the measurements of five lines, and is in almost perfect agreement with the
value derived by Tomkin et al.\ (1993) from a single carbon line. A considerable difference
in the adopted \Teff s between Tomkin el al.\ and our values has little influence probably
because the temperature dependence of the carbon line strengths is weak in the region
from 19\,000 to 24\,000\,K, where \ion{C}{ii} lines reach their maximum strength.
{ A comparison between the renormalised disentangled spectrum of the primary component
and a grid of the theoretical spectra for several \ion{C}{ii}
lines in the 5130--5154\,\AA\ region  is shown in Fig.\,4 (upper panel).

The nitrogen abundance is based on measurements of 17 lines, and is firmly determined
to be $0.20 \pm 0.06$\,dex above the solar value. In turn, this gives the [N/C] abundance
ratio $0.05 \pm 0.03$\,dex, considerably different to the `standard' cosmic ($-0.54 \pm
0.06$; Nieva \& Przybilla 2012) or solar ($-0.61 \pm 0.08$; Asplund et al.\ 2009) values.
Changes in the N/C abundance ratio in the course of mass transfer preserve the imprint
of the components' evolutionary history. These findings are discussed in the next section,
in the context of the chemical evolution in mass transfer binary systems, and provide
a strong argument for case A evolution for the u\,Her binary system.  In Fig.\,4 (middle
panel) the comparison of the three \ion{N}{ii} lines in the 4225--4245\,\AA\ spectral region
to a grid of theoretical spectra are shown. Fig.\,4 (bottom
panel) shows a portion of the spectrum containing \ion{O}{ii} lines.

Helium is a final product of CNO nucleosynthesis and its abundance steadily increases
during the components' evolution.  The helium abundance derived for the primary
component is in perfect agreement with the value found by Nieva \& Przybilla (2012)
for OB stars, albeit that the uncertainty in our determination is quite large. The quality
of the fits for selected \ion{He}{i} lines are shown in Fig.\,5.
The model calculations (Sect.\,6) show an
increase in the helium abundance by mass fraction after the phase of mass transfer
by a factor of approximately 1.25, which settled again to almost the initial value after
thermohaline mixing. The remaining helium enhancement of only 2\% could not be
detected in our measurements because it is below the level of the uncertainties.

The three metals magnesium, silicon and aluminium have a marginally subsolar abundance,
giving on average [M/H] $= -0.04 \pm 0.03$. In our subsequent modelling we therefore
assumed a solar composition.

Despite the importance we did not attempt a detailed abundance analysis for the
secondary star because its renormalised disentangled spectrum suffers from low S/N.
In combination with a high projected rotational velocity, $v\sin i \sim 100$\kms,
this makes the results unreliable. However, we notice a complete absence
of the \ion{C}{ii} 4267\,\AA\ line, which should be visible given the \Teff\ of this star.
 This is illustrated in Fig.\,6 in which \ion{C}{ii}
4267\,\AA\ lines
for both components are shown. The optimal fit for the primary's
line gives the abundance $\log \epsilon(C) = 7.90$ (c.f.\ Table\,5). It is clear
that the primary's carbon abundance does not hold for the secondary. A rough
estimate yields a carbon abundance for the secondary of  $\log \epsilon(C) \leq 7.5$, which is
more than an order of magnitude lower than the `present-day cosmic abundance' of
carbon (Nieva \& Przybilla 2012), also indicated in Fig.\,6. The calculations
to be presented in Sect.\,6 give a depletion of carbon by a factor of
$\sim$7.5 after mass transfer phase has been settled, hence the expected carbon
abundance in the atmosphere of the secondary would be $\log \epsilon(C) \sim 7.4$.
Non-detection of the secondary's \ion{C}{ii} 4267\,\AA\ line therefore corroborates
the predictions of the theoretical models. However, additional
spectra of u\,Her are needed to enhance the S/N of disentangled spectrum of the
secondary star to enable a more definitive conclusion.

\begin{figure*}
\includegraphics[scale=0.75]{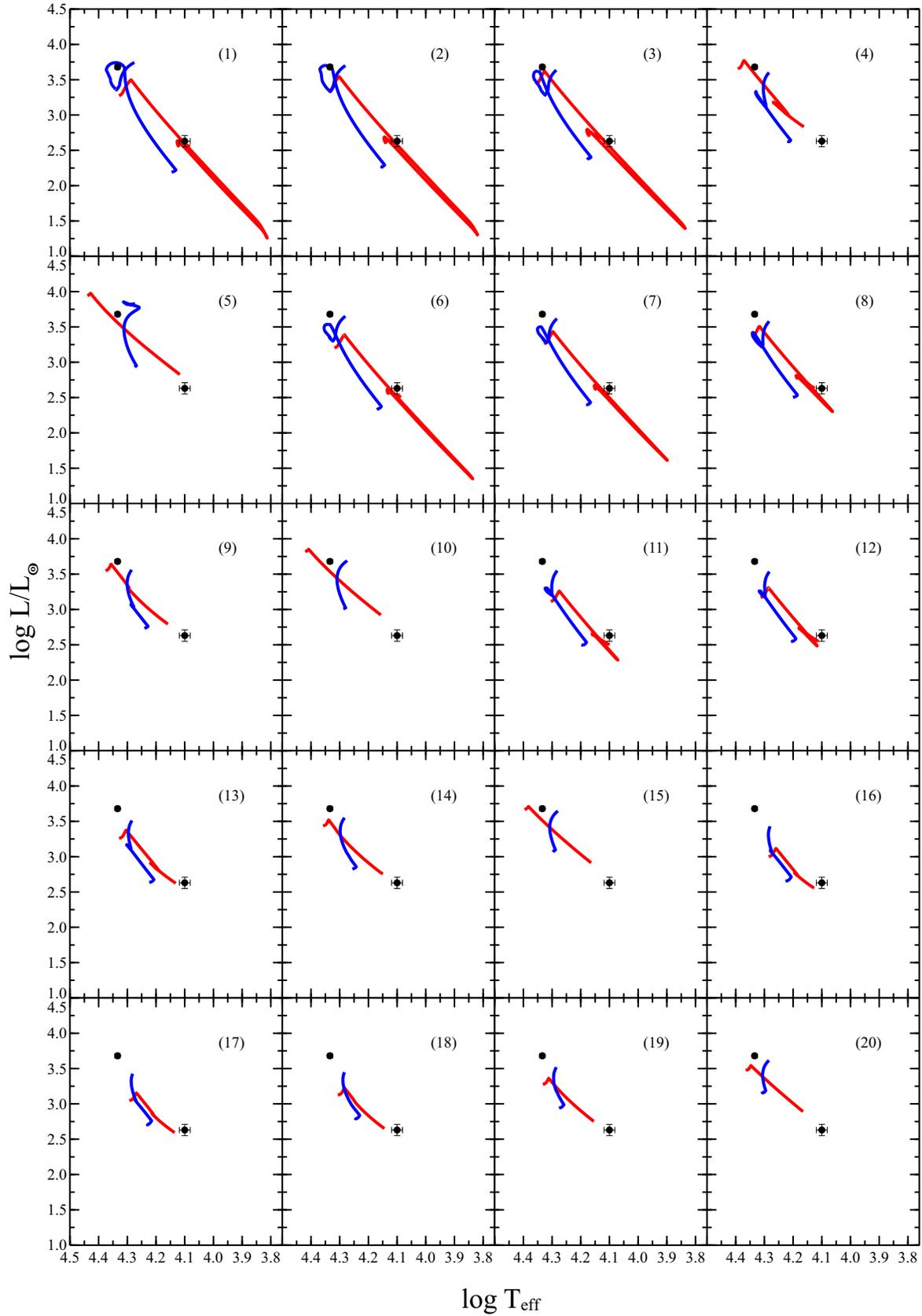}
\caption{The grid of binary evolution tracks on the HR diagram. For given ID from
Table\,6, we show the evolution of the mass donor (red) and mass gainer
(blue), as well as u\,Her's observed $\log L/\Lsun-\log (\Teff$/K). Each set of
tracks was terminated when the mass gainer fills its Roche lobe at the end of SMT.}
\label{tracks}
\end{figure*}

\section{Evolutionary analyses}                                         \label{sec:evo}

As discussed above, u\,Her belongs to a special group of hot Algols, which differ from
ordinary systems by having a larger total mass and mass ratio. Eighty years of accumulated
photometric data show no evidence for an orbital period change, which suggests that
u\,Her is in a very advanced episode of slow mass transfer (SMT). But this finding
is puzzling given its short orbital period, as one would expect a wider orbit at the
end of the mass transfer phase. In any binary, whenever mass transfer occurs from
the higher mass to the lower mass companion, it is expected that the orbital period
decreases until mass ratio reversal. So, tracing back the mass exchange clearly shows
that this system may be in a contact configuration during the rapid mass transfer (RMT)
phase.

It has been considered that u\,Her is a product of case A mass transfer (Webbink 1976).
The first detailed binary evolution models have been done by Nelson \& Eggleton (2001).
Based on a grid of 5500 binary tracks with various values of initial primary mass, mass
ratio and period, they found that the best fitting initial model to produce a system
like u\,Her has an initial donor-star mass of $M_d^i\sim6.31$ \Msun, a mass ratio of
$q^i\sim1.41$ and an orbital period of $P^i\sim1.32$\,d. But they restricted their
calculations to a conservative approach and avoided all contact encounters during
mass transfer. Their result for u\,Her can therefore be taken as a maximum initial
mass and period pair which cannot get into contact during the RMT phase.

de Mink et al.\ (2007) extended this study with 20\,000 detailed calculations of
binary evolutionary tracks using a modified code based on that of Nelson \& Eggleton
(2001).  They modified the Nelson \& Eggleton code so that stellar
structure equations of both stars are solved simultaneously, which is needed for
accurately modelling mass transfer phases. Moreover, they
accounted for non-conservative mass loss and short contact phases during
RMT, and concentrated on binaries in the Small Magellanic Cloud. de Mink et al.\ (2007)
suggested a new subtype of Algol systems (AR (rapid contact) $\rightarrow$ AN (no contact))
which shows a short-lived
 contact phase during the thermal response of the mass gainer to the RMT. After this the
 mass gainer restores its thermal equilibrium and shrinks, then mass transfer proceeds.
To evaluate non-conservative evolution they introduced a mass transfer efficiency parameter
($\beta$) which is a measure of how much matter is lost relative to that transferred. For
the angular momentum evolution, they assumed that all the matter is lost via bi-polar
emission from the mass-accreting star hence carries this star's specific angular momentum.
One of the hot Algols in their sample (OGLE 09\,064498) has a configuration very close to
that of u\,Her: $M_p\sim8.4\pm0.7$\Msun, $q\sim0.323$ and $P\sim2.64$\,d. The best-fitting
initial model that they found had $M_d^i\sim7.10$\Msun, $q^i\sim1.68$ and $P^i\sim1.34$\,d.

Instead of making a large grid of binary tracks which is more suited for a large sample
of systems, we specifically prepared initial models for u\,Her for different initial mass
ratios and mass loss rates. Since the uncertainty on angular momentum loss has a big
influence on our understanding of binary evolution, there is little value in using
a very fine grid in parameter space. We then made some simplifications to reduce the
number of initial models to reach plausible results. After determining the initial
parameters, we searched for the best fitting models produced by the binary evolution
code in this grid.

We first considered four main sets of initial mass ratios $q^i$: 1.25, 1.50, 1.75 and
2.00. These are typical values to produce Algol-like systems at the end of mass transfer.
We did not go beyond $q^i>2.0$ because mass-gaining stars in these systems are are unlikely
to regain thermal equilibrium during RMT. To prepare a subset of initial models to take
into account non-conservative mass transfer, we adopted the approach of de Mink et al.\
(2007). The mass transfer efficiency parameter $\beta$ is defined as,
\begin{equation} \label{beta}
\beta=1-\vert\frac{\dot{M_g}}{\dot{M_d}}\vert \qquad 0\leq\beta\leq1
\end{equation}
where $g$ denotes the mass gainer and $d$ the mass donor. From Eq.\,\ref{beta}, one can
easily see that $\beta=0$ corresponds to conservative evolution. To evaluate angular
momentum loss, we used the Hurley et al.\ (2002) approximation which assumes that mass
loss takes the specific angular momentum of the mass-losing star. This is likely true
for case A evolution due to the lack of an accretion disc producing bipolar mass loss.
Using this approximation and taking logarithmic differentiation of the angular momentum
equation for a two-mass system
\begin{equation}
J^2=(G\frac{M_d^2M_g^2}{M_d+M_g})4\pi^2\,A
\end{equation}
where $A$ is a separation of binary, one can easily derive a relation for the orbital
period evolution with the help of Kepler's second law:
\begin{align}\label{per}
\frac{P^f}{P^i} &= &\left( \frac{M_d^i+M_g^i}{M_d^f+M_g^f}\right)^{1/2}
\left( \frac{M_g^i}{M_g^i+(1-\beta)(M_d^i-M_d^f)}\right)^3 \nonumber \\
&&\left( \frac{M_g^i+M_d^i}{M_g^i+(1-\beta)M_d^i+\beta M_d^f}\right)^{-3/2}
\left( \frac{M_d^i}{M_d^f}\right)^{3(1-\beta)}
\end{align}
where superscripts $i$ and $f$ stand for initial and final parameters. The evolution
of the total mass of system is adapted from Giuricin \& Mardirossian (1981);
\begin{equation} \label{mass}
\frac{M_t^i}{M_t^f}=\frac{(1+q^i)}{(1+q^f)}\frac{[1+q^f(1-\beta)]}{[1+q^i(1-\beta)]}.
\end{equation}

Using a range of $\beta =  [0.0, 0.10, 0.25, 0.50, 0.75]$, i.e.\ from conservative to highly
non-conservative, we created 20 different initial models as candidate progenitors of u\,Her
($M_g^f\sim7.9\pm0.26$, $q^f\sim0.35\pm0.02$ and $P^f=2.05$\,d). But since Eqs.\,\ref{per}
and \ref{mass} do not consider the properties of the stellar structure under the effect
of mass transfer, one has to run detailed evolution codes to compare all of the observed
properties of each companion as well as the orbit.

To calculate detailed binary evolution tracks, we used the Cambridge version of the
{\sc stars}\footnote{Freely avaliable at
{\tt http://www.ast.cam.ac.uk/\textasciitilde stars/}} code which was originally
developed by Eggleton (1971, 1972). The most recent updates allow calculation of
the evolution of each component simultaneously, the prescription of mass transfer
and various physical improvements, and are explained in Stancliffe \& Eldridge (2009).
Since both observed and candidate initial masses of the components are in the
intermediate-mass regime, we fixed the overshooting parameter at $\delta_{os} = 0.12$.
We also assumed a solar composition in all of our components at the ZAMS. Each binary
evolution track was terminated whenever the mass gainer filled its Roche lobe at
the end of SMT, i.e.\ reverse mass transfer.

\begin{figure}
\centering
\includegraphics[width=7.4cm, angle=0]{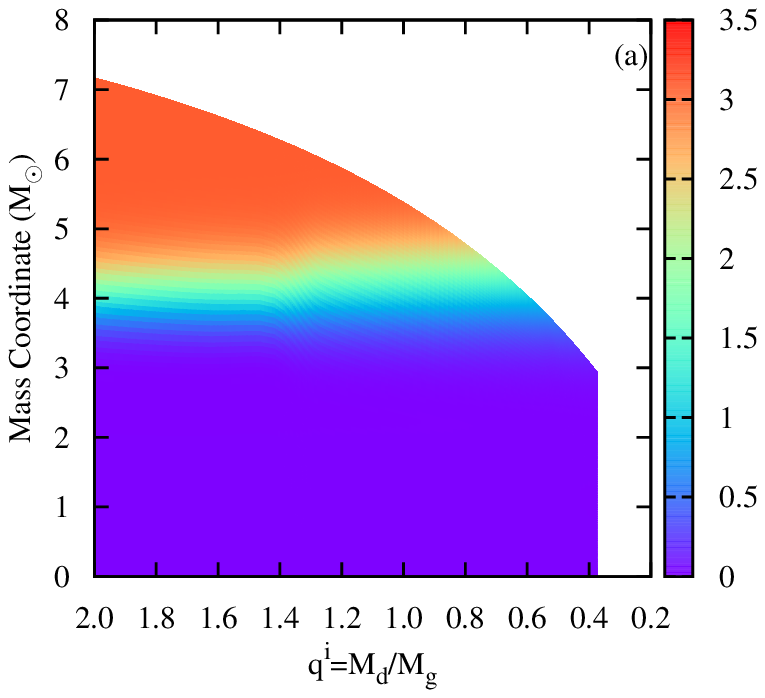} \\
\includegraphics[width=7.4cm, angle=0]{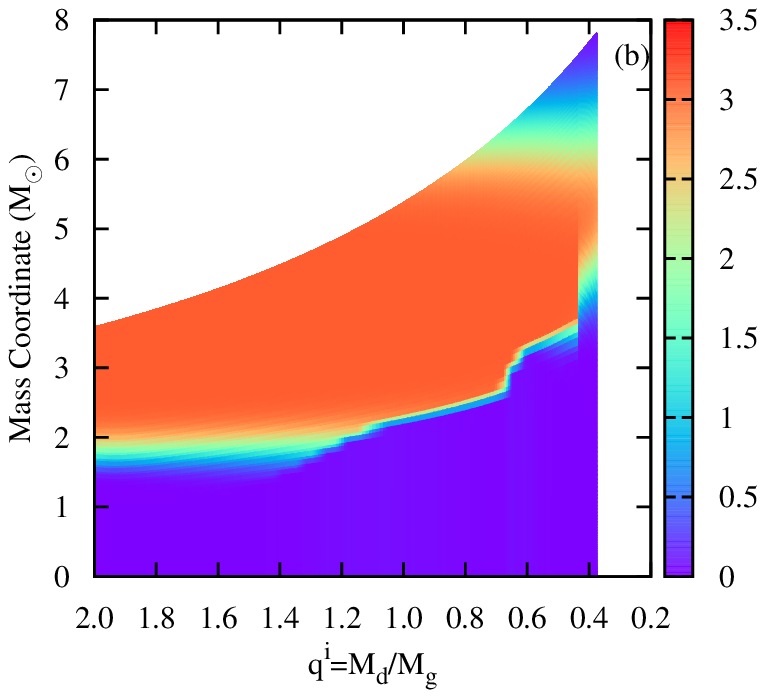} \\
\includegraphics[width=7.4cm, angle=0]{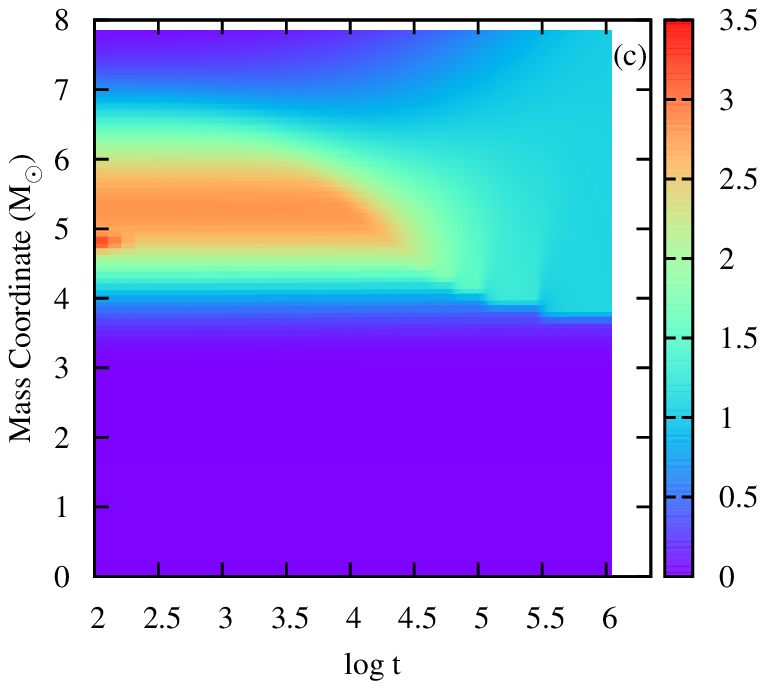}
\caption{The internal C/N ratio changes at mass coordinates for the mass donor (panel 1) and
the mass gainer (panel 2) during the mass ratio change as an indicator of rapid mass transfer
 (RMT),
derived from the evolution tracks of the best fitting model -- ID(1). The effect of
thermohaline convection on the internal profile of mass gainer after  slow mass transfer (SMT),
$t= 0$ and $q^i < 0.4$} is shown in  panel 3.
\label{cn}
\end{figure}

\begin{table*}
\caption{The list of initial and final parameters for our binary evolution grid.
The first eight columns are input parameters of {\sc stars} evolution code runs.
The remainder are the best fitting model results for representing the current
state of u\,Her. Run ID(1) is chosen to be the best fitting progenitor system.}
\begin{tabular}{rcccccccccccccccc}
\hline
ID	&$q^i$	&$\beta$	&$M_t^i$	&$M_g^i$	&$M_d^i$	&$P_{\rm lim}$	&$P^i$
&$M_d^f$	&$M_g^f$	&$P^f$	&$\log L_d$	&$\log T_d$	&$\log R_d$	&$\log L_g$	&$\log T_g$	&$\log R_g$\\
 	& 	& 	&\Msun 	&\Msun &\Msun 	&d 	&d 	&\Msun &\Msun 	&d 	&\Lsun 	&K 	&\Rsun
&\Lsun 	&K 	&\Rsun \\
\hline
1	&2.00	&0.00	&10.74	&3.58	&7.16	&0.77	&1.35	&2.90	&7.84	&1.93	&2.56	&4.09	&0.63	&3.73	&4.28	&0.83\\
2	&2.00	&0.10	&11.21	&3.74	&7.47	&0.78	&1.29	&3.06	&7.71	&1.77	&2.60	&4.11	&0.62	&3.69	&4.28	&0.80\\
3	&2.00	&0.25	&12.04	&4.01	&8.03	&0.80	&1.20	&3.37	&7.51	&1.59	&2.68	&4.13	&0.60	&3.62	&4.29	&0.76\\
4	&2.00	&0.50	&14.00	&4.67	&9.33	&0.84	&1.06	&3.86	&7.41	&1.54	&2.85	&4.17	&0.61	&3.58	&4.29	&0.74\\
5	&2.00	&0.75	&17.26	&5.75	&11.51	&0.90	&0.91	&3.16	&7.83	&2.34	&2.84	&4.12	&0.70	&3.83	&4.28	&0.88\\
6	&1.75	&0.00	&10.74	&3.91	&6.83	&0.76	&1.19	&3.16	&7.58	&1.65	&2.51	&4.09	&0.60	&3.64	&4.28	&0.77\\
7	&1.75	&0.10	&11.17	&4.06	&7.11	&0.77	&1.15	&3.31	&7.48	&1.55	&2.56	&4.11	&0.59	&3.60	&4.29	&0.75\\
8	&1.75	&0.25	&11.94	&4.34	&7.60	&0.79	&1.10	&3.56	&7.37	&1.48	&2.64	&4.13	&0.59	&3.56	&4.29	&0.73\\
9	&1.75	&0.50	&13.69	&4.98	&8.71	&0.82	&1.00	&4.03	&7.34	&1.45	&2.81	&4.16	&0.60	&3.54	&4.29	&0.72\\
10	&1.75	&0.75	&16.51	&6.01	&10.51	&0.87	&0.90	&3.85	&7.67	&1.89	&2.94	&4.16	&0.67	&3.68	&4.28	&0.81\\
11	&1.50	&0.00	&10.74	&4.30	&6.44	&0.75	&1.07	&3.46	&7.28	&1.42	&2.51	&4.10	&0.57	&3.53	&4.29	&0.72\\
12	&1.50	&0.10	&11.13	&4.45	&6.68	&0.76	&1.05	&3.57	&7.25	&1.39	&2.56	&4.12	&0.57	&3.52	&4.29	&0.71\\
13	&1.50	&0.25	&11.81	&4.72	&7.09	&0.77	&1.01	&3.80	&7.19	&1.35	&2.64	&4.14	&0.57	&3.49	&4.29	&0.70\\
14	&1.50	&0.50	&13.34	&5.33	&8.00	&0.80	&0.96	&4.02	&7.32	&1.44	&2.77	&4.15	&0.60	&3.53	&4.29	&0.72\\
15	&1.50	&0.75	&15.70	&6.28	&9.42	&0.84	&0.91	&4.12	&7.60	&1.71	&2.93	&4.17	&0.65	&3.63	&4.28	&0.77\\
16	&1.25	&0.00	&10.74	&4.77	&5.97	&0.73	&0.98	&3.82	&6.92	&1.22	&2.57	&4.13	&0.54	&3.41	&4.28	&0.66\\
17	&1.25	&0.10	&11.07	&4.92	&6.15	&0.74	&0.97	&3.93	&6.92	&1.22	&2.61	&4.14	&0.55	&3.41	&4.28	&0.66\\
18	&1.25	&0.25	&11.66	&5.18	&6.48	&0.75	&0.96	&4.05	&7.00	&1.25	&2.66	&4.15	&0.56	&3.43	&4.29	&0.67\\
19	&1.25	&0.50	&12.93	&5.74	&7.18	&0.77	&0.95	&4.18	&7.24	&1.36	&2.77	&4.16	&0.59	&3.50	&4.29	&0.70\\
20	&1.25	&0.75	&14.80	&6.58	&8.22	&0.81	&0.95	&4.31	&7.56	&1.58	&2.91	&4.17	&0.64	&3.60	&4.29	&0.75\\
\hline
\end{tabular}
\label{tab:in}
\end{table*}

In Table\,6, we list the grid of our binary tracks. We show the initial
parameters of the systems as well as the best fitting model compared to the observed
absolute parameters in Table\,2. We checked each system's initial
period with limiting period, i.e.\ the smallest period for given binary, via this
equation from Nelson \& Eggleton (2001):
\begin{equation}
P_{\rm lim}\approx\frac{0.19M_d^i+0.47{M_d^i}^{2.33}}{1+1.18{M_d^i}^2}.
\end{equation}

We also show the binary tracks and observed parameters of the system on the HR diagram
in Fig.\,7. Providing that each system starts with different initial parameters,
the thermal responses of each component determine the length of the RMT and SMT phase.
Most of the systems in Table\,6 cannot accrete enough mass to reach the
observed masses of the components of u\,Her before reverse mass transfer. Based on
$\chi^2$ minimisation and visual inspection, the best fitting model belongs to a group
of conservative and high initial mass ratio systems. This was also the case for OGLE
09\,064498 as discussed above. We also noticed a short-term contact phase, as also
discussed by de Mink et al.\ (2007), in high initial mass ratio systems for the case
of highly efficient mass transfer $q^i\geq1.75$ and $\beta\leq0.25$.

Finding the best initial model parameters allowed us to trace the chemical evolution
of both components during mass transfer. In Fig.\,8a,b we show the change
in internal profile of the ratio C/N from the centre to the surface of each component.
Due to the very different timescales of the RMT and SMT phases, we plot this change
versus the mass ratio instead of time. One can easily recognize the abrupt internal
change in the mass gainer's profile which corresponds to the transition from RMT to
SMT. From Fig.\,8a, we find that the mass donor lost its mass up to the depth at
which the CNO cycle reduced the C/N ratio from the cosmic ($\sim$3.2) to the equilibrium
($\sim$0.1) value. This nucleosynthetically processed material was then accreted on the
surface of mass gainer. The accreted material had a higher mean molecular weight than
that from the surface of the mass gainer. In such a case, one may expect thermohaline
convection to mix this material and alter the surface composition. As shown by
Stancliffe \& Eldridge (2009), the effect of thermohaline mixing on the surface
is negligible during RMT. We therefore ran all of our tracks without thermohaline
mixing to find the lower limit of the C/N ratio on the surface. We then applied
thermohaline mixing to the model of the mass gainer to trace the change of chemical
composition on its surface. As the thermohaline condition is not satisfied, we ignored
 the mass donor.

In Fig.\,8c, we show the effect of thermohaline mixing on the whole internal
profile of the star. Due to the material originating from different layers of mass
donor, the outer layers of the stars have a variable composition profile. We find
that the thermohaline mixing alters the surface composition of the stars on relatively
short timescale ($\sim$$10^5$\,yr). Thus we expect the surface C/N ratio of the gainer
to be between non-mixed ($\sim$$0.1$) to mixed ($\sim$$1$). This result is in good
agreement with our observed ratio of C/N $= 0.89$.

We believe that determining the composition of the mass donor would be an important
opportunity to constrain the initial evolutionary parameters. Such a situation would
allow us to build a fine grid of binary tracks to compare results with observations
as well as our understanding of the processes involved in binary evolution such as
mass loss mechanisms and thermohaline mixing. Even though we could not determine
the surface composition of the mass donor, the lack of a prominent \ion{C}{ii}
4267\,\AA\ line compared to single stars of the same \Teff\ is a strong indication
 of decreased carbon abundance on the surface as a result of case A mass transfer.
This is because, in a wider orbit, the mass donor may only lose the upper layers
without reaching CNO processed regions. So far our evolutionary calculation shows
that u\,Her could start its evolution with $M_d^i\sim7.16$\Msun, $q^i\sim2.00$ and
$P^i\sim1.35$\,d.

\section{Conclusions}

Large-scale mass transfer in Algol-type binary systems not only leads to mass reversal
and an exchange of the role between the components but also affects the photospheric
chemical composition (Sarna 1992). Formerly deep layers can become exposed after
a short-lived mass exchange. Therefore tracing the photospheric elemental abundances
of the components might help to constrain their past. Already in the first observational
 studies of this effect, a clear evidence for carbon underabundance in the (brighter)
primaries of Algol-type systems was found (Parthasaranty et al. 1983, Cugier \& Hardorp
1988, Cugier 1989, Tomkin et al.\ 1993). More recent studies have not challenged that
general conclusion (Glazunova et al.\ 2011, \.{I}bano\v{g}lu  et al.\ 2012).

The technique of spectral disentangling has opened up new possibilities in detailed
spectroscopic studies of the stars in binary or multiple systems (c.f.\ Pavlovski
\& Hensberge 2005, Pavlovski \& Southworth 2009) since it enables isolation of the
individual spectra of the components while simultaneously solving for the orbital
elements. This allows a detailed determination of the photospheric chemical composition
to be put onto much firmer ground.

This work addresses the chemical evolution of u\,Her, a hot Algol-type binary system.
In hot Algols both components are of spectral type B, and hence more massive than in
a typical Algol system. Consequently, the CNO cycle is more efficient and a greater
alteration of the chemical composition is expected. A new set of 43 high-resolution
\'echelle spectra were secured and analysed using spectral disentangling. Our conclusions
are:

\begin{itemize}

\item The masses and radii of the components are $M_{\rm A} = 7.88 \pm0.26$\Msun\ and
$R_{\rm A} = 4.93\pm 0.15$\Rsun\ for the primary (mass-gaining) star, and $M_{\rm B} =
2.79\pm0.12$\Msun\ and $R_{\rm B} = 4.26 \pm 0.06$\Rsun\ for the secondary (mass-losing)
star.

\item The \Teff s were determined from an optimal fitting of the H$\gamma$ and H$\beta$
lines in the stars' disentangled and renormalised spectra. The light ratio was determined
from re-analysis of the {\it Hipparcos photometry}. We find $T_{\rm eff,1} = 21\,600\pm220$\,K
 and $T_{\rm eff,2} = 12\,600\pm550$\,K.

\item The primary star's photospheric elemental abundances were derived from an extensive
line list in the whole optical spectrum and exhibit a carbon depletion of [C/H] $= -0.47
\pm 0.05$ and a nitrogen enhancement of [N/H] $= 0.20 \pm 0.06$ with respect to the
standard cosmic abundance pattern (Nieva \& Przybilla 2012). The uncertainty in the
helium abundance does not allow a firm conclusion on the possible enhancement of the
helium abundance ([He/H] $= 0.02 \pm 0.05$).

\item Theoretical evolutionary calculations reproduce the current characteristics of
the system from a progenitor binary with intial masses $M_{\rm d} \sim 7.2$\Msun\ and
$M_{\rm g} \sim 3.6$\Msun, and an initial period $P_{\rm i} \sim 1.35$\,d. The calculations
have shown that thermohaline mixing alters the surface composition of stars on a relatively
short timescale, $t \sim 10^5$\,yr. The observed C/N abundance ratio ($\sim$$0.9$)
corroborates this picture and indicates a strong mixing of the stellar material.

\item The composition of the secondary component would be a further
important constraint on the initial properties of the u\,Her system, but requires spectra
of a higher signal to noise ratio. A non-detection of the \ion{C}{ii} 4267\,\AA\ line
corroborates the model calculations and a general picture emerged from the
study of the primary's photospheric chemical composition.

\end{itemize}

\section*{Acknowledgements}

We thank the referee for constructive and timely comments.
Based on observations collected at the Centro Astron\'omico Hispano Alem\'an (CAHA) at Calar Alto,
operated jointly by the Max-Planck Institut f\"ur Astronomie and the Instituto de Astrof\'{\i}sica
de Andaluc\'{\i}a (CSIC) through the OPTICON observing program.
VK was supported by a Croatian MZOS PhD grant. KP acknowledges funding from  the  University
of Zagreb research grant. AD was supported by the Turkish Scientific and Technical Research
Council (T\"{U}B\.{I}TAK) research grant 113F067. JS acknowledges financial support from STFC in the form of an Advanced Fellowship.


\end{document}